\documentclass[review]{elsarticle}

\usepackage{multirow,setspace,amssymb,amsmath,graphicx,color,rotating,subfigure,url}
\usepackage{natbib}
\usepackage{textcomp}
\usepackage{CJK}
\usepackage{bm}%
\usepackage{rotating}
\usepackage{epsfig}% Include figure files
\usepackage{dcolumn}% Align table columns on decimal point
\usepackage{epstopdf}

\usepackage{hyperref}
\usepackage{graphicx}% Include figure files
\usepackage{dcolumn}% Align table columns on decimal point
\usepackage{bm}% bold math
\usepackage{color}% bold math
\usepackage{amsmath}
\usepackage[utf8]{inputenc}
\usepackage[T1]{fontenc}
\usepackage{amssymb}
\usepackage{enumerate}
\usepackage[superscript, biblabel, nomove]{cite}
\usepackage[ruled,linesnumbered]{algorithm2e}
\usepackage{amsmath}
\usepackage{booktabs}
\usepackage[ruled]{algorithm2e}
\newtheorem{myDef}{Definition}
\usepackage{indentfirst}

\begin{document}
\begin{frontmatter}

\title{Influence Maximization in Hypergraphs}

\author[inst1]{Ming Xie}
\author[inst1]{Xiu-Xiu Zhan}\ead{zhanxiuxiu@hznu.edu.cn}  %% Corresponding author
\author[inst1]{Chuang Liu}\ead{liuchuang@hznu.edu.cn}
\author[inst2]{Zi-Ke Zhang}\ead{zkz@zju.edu.cn}  %% Corresponding author

\address[inst1]{Research Center for Complexity Sciences, Hangzhou Normal University, Hangzhou, 311121, P. R. China}
\address[inst2]{College of Media and International Culture, Zhejiang University, Hangzhou 310058, P. R. China}

\begin{abstract}
Influence maximization in complex networks, i.e., maximizing the size of influenced nodes via selecting $K$ seed nodes for a given spreading process, has attracted great attention in recent years. However, the influence maximization problem in hypergraphs, in which the hyperedges are leveraged to represent the interactions among more than two nodes, is still an open question. In this paper, we  propose an adaptive degree-based heuristic algorithm, i.e., \emph{Heuristic Degree Discount} (\emph{HDD}), which iteratively selects nodes with low influence overlap as seeds, to solve the influence maximization problem in hypergraphs. We further extend algorithms from ordinary networks as baselines and compare the performance of the proposed algorithm and baselines  on both real data and synthetic hypergraphs. Results show that HDD outperforms the baselines in terms of both effectiveness and efficiency. Moreover, the experiments on synthetic hypergraphs indicate that HDD shows high performance especially in hypergraphs with heterogeneous degree distribution.
\end{abstract}


\end{frontmatter}

\makeatletter
\def\ps@pprintTitle{%
  \let\@oddhead\@empty
  \let\@evenhead\@empty
  \let\@oddfoot\@empty
  \let\@evenfoot\@oddfoot
}
\makeatother
% \linenumbers
\section{Introduction}
\makeatletter
\newcommand{\rmnum}[1]{\romannumeral #1}
\newcommand{\Rmnum}[1]{\expandafter\@slowromancap\romannumeral #1@}
\makeatother

% \section*{INTRODUCTION}
 As a classical optimization problem, influence maximization aims to find a  set of $K$ initial spreaders that maximize the influence spread under a certain spreading dynamics in a network. Due to its abundant applications, e.g., the control of disease ~\cite{cheng2020outbreak}, the dissemination of information ~\cite{zhang2014recent,lei2015online} and viral marketing~\cite{domingos2001mining,chen2010scalable}, the problem is widely studied in recent years. Extensive researches of influence maximization are oriented to ordinary networks, where nodes represent individuals and edges represent pairwise interactions between individuals. The problem was first proposed to identify the most helpful customers in viral marketing~\cite{domingos2001mining}. Later on, Kempe et al.~\cite{kempe2003maximizing} used two popular diffusion models, Independent Cascade model and Linear Threshold model, in influence maximization. They modeled this problem as a combinatorial optimization problem, which is proved to be NP-hard, and proposed a greedy algorithm which can guarantee a $(1-\frac{1}{e}-\epsilon)$ approximation ratio for selecting the seed nodes to tackle it.
In addition, the cost effective lazy forward method (CELF~\cite{leskovec2007cost}) and its improved variant (CELF++~\cite{goyal2011celf++}) were designed respectively to improve the efficiency of the greedy algorithm. Moreover, there are many other methods proposed to improve the efficiency and the accuracy of influence maximization\cite{goyal2011simpath, chen2014cim, nguyen2016stop, narayanam2010shapley}, including maximum influence
arborescence (MIA)~\cite{chen2010scalable}, Prefix excluding MIA (PMIA)~\cite{wang2012scalable} and Two-phase Influence Maximization (TIM)~\cite{tang2014influence}.

In many real-world scenarios, an edge in ordinary networks with dyadic relationship could hardly characterize the interactions if the interactions involve more than two entities. For example, many users could form groups for information sharing in social platforms, more than two researchers may contribute to one scientific paper, and many people might be listed in mass emails. This kind of relations could be represented by a hypergraph~\cite{cencetti2021temporal, young2021hypergraph, kitsak2020link}, with hyperedges characterizing the  polyadic interactions
among more than two nodes~\cite{ouvrard2020hypergraphs}. In light of influence maximization in hypergraphs, it is still a mostly unexplored problem with only a few studies focusing on this field.
For instance, Amato et al.~\cite{amato2017influence} modelled the social media network via a hypergraph, in which user-to-multimedia relationships  are represented by hyperedges.
They further applied algorithms, such as TIM+~\cite{tang2014influence} and IMM~\cite{tang2015influence}, which were proposed to solve influence maximization problem in bipartite graphs, to tackle the influence maximization in a hypergraph after transforming it to a bipartite graph.
Zhu et al.\cite{zhu2018social} proved that influence maximization in directed hypergraphs under independent cascade model is a NP-hard problem and designed a sandwich framework which provides a $(1-\frac{1}{e}-\epsilon)$  approximation ratio with high computational complexity to solve it. A ranking-based algorithm was proposed to solve the influence maximization problem under the HyperCascade model \cite{ma2022hyper}, where the model actually considers spreading process on the bipartite augment graph of a hypergraph. In addition, a set of greedy-based heuristic strategies were proposed to solve the minimum target set selection problem in hypergraphs~\cite{antelmi2021social}. However, the above researches either considered to transform hypergraphs to bipartite graphs or designed greedy algorithms  to solve the influence maximization problem in hypergraphs,  ignoring the basic hypergraph topological structures which may play a crucial role  in solving the influence maximization problem. In this paper, we aim to explore how to utilize the basic topological properties of a hypergraph for influence maximization.

Degree centrality, as an essential topological property, was frequently used to characterize the node importance in a network~\cite{lu2016vital, stegehuis2021network}. In this paper, we address the problem of how to choose the initial seeds  for influence maximization in hypergraphs based on the node degree. Firstly, we design a discrete-time susceptible-infected (SI) model with Contact Process (CP)~\cite{suo2018information} to
model the influence spreading process in hypergraphs. Then, we propose Heuristic Degree Discount (HDD) algorithm for influence maximization, which iteratively avoids choosing nodes that have large influence overlap with the existing seeds as the seed candidates. Experiments indicate that the proposed algorithm can achieve better performance efficiently and accurately compared with the baselines on both real-world data and synthetic hypergraphs.

The remainder of this paper is organized as follows. Section 2 presents preliminary definitions of a hypergraph and the description of real-world data. In Section 3, we illustrate the influence maximization problem in hypergraphs, the spreading model as well as the algorithms. Experimental results and analysis are given in Section 4. Finally, we conclude the paper in Section 5.

\section{Representation of a Hypergraph}
\subsection{Definition of a Hypergraph}
\begin{figure}[h!]
\centering		
\includegraphics[scale=1.2]{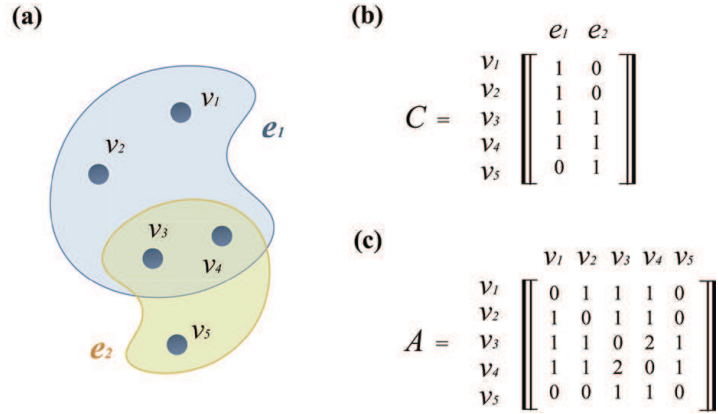}	
\caption{An illustration example of (a) a hypergraph with 5 nodes and 2 hyperedges; (b) the incidence matrix of (a); (c) the adjacency matrix of (a).}
\label{fig:hypergraph ex}		
\end{figure}

A hypergraph is represented as $H(V,E)$, in which  $V=\{ v_1, v_2, ..., v_n \}$ is the node set and $E = \{ e_1, e_2, ..., e_m \}$ is the hyperedge set. An incidence matrix of $H$ is given by $C=(c_{i\alpha})_{[n\times m]}$,  where
\begin{equation}
    c_{i \alpha}=
   \begin{cases}
   1&\mbox{if $v_i \in e_{\alpha}$}\\
   0&\mbox{otherwise}
   \end{cases}
  \end{equation}
Therefore, the adjacency matrix $A_{[n\times n]}$ can be derived from $C$,
\begin{equation}
    A_{ij} = [CC^T - D]_{ij},
  \end{equation}
where $D$ is a diagonal matrix whose diagonal elements represent the number of hyperedges a node belongs to. The value of $A_{ij}$ indicates the number of hyperedges that contain both node $v_i$ and node $v_j$. An example of a hypergraph is given in Figure~\ref{fig:hypergraph ex}, which contains 5 nodes and 2 hyperedges. The incidence matrix $C$ and adjacency matrix $A$ are also given correspondingly.

Given the incidence matrix of a hypergraph, we can define the degree of a node in a hypergraph in two different ways~\cite{battiston2020networks}, i.e., the degree and the hyperdegree. The degree of a node $v_i$ ($deg(i)$) indicates the number of neighboring nodes that $v_i$ is adjacent to, which can be expressed as follows:

\begin{equation}
deg(i) = \sum_{j=1}^{n}A^{(b)}_{ij},
\end{equation}

where
\begin{equation}
A^{(b)}_{ij}=\left\{
	\begin{aligned}
	1 \quad if~~A_{ij} > 0\\
	0 \quad if~~A_{ij} = 0\\
	\end{aligned}
	\right
	.
\end{equation}

The hyperdegree of node $v_i$ is defined as the number of hyperedges that node $v_i$ belongs to, which is given by the following equation:
\begin{equation}
    d^{H}(i) = \sum_{j=1}^{m}C_{ij}
  \end{equation}

According to the above definitions, we can calculate the degree and hyperdegree of the nodes in Figure~\ref{fig:hypergraph ex}. For instance, the degree of node  $v_3$ is $deg(3)=4$ and the hyperdegree of node  $v_3$ is $d^{H}(3)=2$.

\subsection{Data description}
We show the basic description and properties of eight hypergraphs generated by real-world datasets, which are collected from different domains\footnote{\url{https://www.cs.cornell.edu/~arb/data/}} \footnote{\url{http://bigg.ucsd.edu/}}. The hypergraphs will be used to evaluate the performance of our algorithms in the subsequent sections. The topological properties of them are given in Table~\ref{tab:1}. The detailed description of each data is given as follows:

 \textbf{cat-edge-algebra-questions dataset (Algebra) \& cat-edge-geometry-questions dataset (Geometry).} The two datasets contain interactions between users on the stack exchange web site Math Overflow. The interactions between users are mainly about comments, questions and answers on algebra (or geometry) problems. Each node represents a user on MathOverflow and hyperedges are sets of users who answered a certain question category (algebra or geometry).

\textbf{cat-edge-madison-restaurant-reviews (Restaurant-Rev).}
    The data indicates users who reviewed an establishment of a particular category (different types of restaurants in Madison, WI) within a month timeframe. Each node represents a user on Yelp and a hyperedge represents a set of users who reviewed a certain restaurant.

 \textbf{cat-edge-music-blues-reviews (Music-Rev).} The data contains nodes representing users on Amazon, and hyperedges are sets of reviewers who reviewed a certain product category (different types of blues music) within a month timeframe.

\textbf{cat-edge-vegas-bars-reviews (Bars-Rev).}
     Each node in the dataset represents a user on Yelp, and a hyperedge is a set of users who reviewed a certain bar in Las Vegas, NV.

 \textbf{NDC-classes.}
    The dataset contains nodes representing class labels, and a hyperedge is a drug which consists of a set of class labels.

\textbf{iAF1260b.}
    The data contains nodes representing reaction-based metabolics, and hyperedges are sets of metabolics which are applied to a certain reaction. The duplicate hyperedges are removed.

\textbf{iJO1366.} Similar to \textbf{iAF1260b}, this is also a metabolic hypergraph with each node representing a reaction-based metabolic, and hyperedges are sets of metabolics which are applied to a certain reaction. The duplicate hyperedges are removed.

 \begin{table}[htb]
    \centering
        \caption{Basic properties of the datasets. $n$ and $m$ are the number of nodes and hyperedges in a hypergraph, $\left\langle deg \right\rangle$ is the average of node degree, $\left\langle d^H \right\rangle$ is the average of node hyperdegree,  $\left\langle d^E \right\rangle$ represents the average of the size of the hyperedges, the size of a hyperedge is given by the number of nodes in the hyperedge. $C$, $\left\langle d \right\rangle$, $dia$ and $ld$ are the clustering coefficient, average shortest path length, diameter and link density of the corresponding ordinary network of a hypergraph.  }
        \resizebox{\textwidth}{!}{
    \begin{tabular}{cccccccccc}
         \toprule
		Data &$n$&$m$& $\left\langle deg \right\rangle$ & $\left\langle d^H \right\rangle$&$\left\langle d^E \right\rangle$&$C$&$\left\langle d \right\rangle$&$dia$&$ld$\\
	\midrule
		Algebra & 423 & 1268 & 78.90 & 19.53 & 6.52 & 0.79 & 1.95 & 5 & 0.19 \\
		Restaurant-Rev & 565 & 601& 79.75 & 8.14 & 7.66 & 0.54 & 1.98 & 5 & 0.14 \\
		Geometry & 580 & 1193& 164.79 & 21.53 & 10.47 & 0.82 & 1.75 & 4 & 0.28 \\
		Music-Rev & 1106 & 694 & 167.87 & 9.49 & 15.13 & 0.62 & 1.99 & 8 & 0.15 \\
		NDC-classes & 1161 & 1088 & 10.71 & 5.55 & 5.92 & 0.61 & 3.50 & 9 & 0.01 \\
		Bars-Rev & 1234 & 1194 & 174.30 & 9.61 & 9.93 & 0.58 & 2.10 & 6 & 0.14 \\
		iAF1260b & 1668 & 2351 & 13.26 & 5.46 & 3.87 & 0.55 & 2.67 & 7 & 0.007 \\
		iJO1366 & 1805 & 2546 & 16.91 & 5.55 & 3.94 & 0.58 & 2.62 & 7 & 0.009 \\
	\bottomrule
    \end{tabular}
        }
    \label{tab:1}
\end{table}

\section{Models and Algorithms}\label{MODELS AND ALGORITHMS}
\subsection{Problem statement}
% \subsection*{Influence maximization problem in hypergraphs}
Given a specific spreading model, the influence maximization problem\cite{li2018influence} aims to identify $K$ spreaders (also called seed nodes) in a network that can maximize the expected number of influenced nodes. Mathematically, the problem can be described as follows,
\begin{equation}\label{IM}
\begin{split}
  argmax\left\{\sigma(S)\right\},\\
  s.t.|S| = K.
\end{split}
\end{equation}
where $\sigma(S)$ represents the expected influence of the seed node set $S$, and the number of nodes $K$ in the seed set is the constraint condition of this problem.

 It has been shown that the influence maximization problem in ordinary networks is NP-hard~\cite{kempe2003maximizing}. And the influence maximization problem in hypergraphs, which can be considered as the generalization of influence maximization in ordinary networks, is also NP hard~\cite{zhu2018social}. That is to say, it cannot be solved in polynomial time. Therefore, we propose to use greedy and heuristic algorithms to approximate its optimal solution.

\subsection{SI spreading model with Contact Process dynamics}
To quantify the spreading influence of the seed nodes~\cite{ferraz2021phase, de2020social}, we propose to use a Susceptible-Infected (SI) model with Contact Process (CP) dynamics on a hypergraph~\cite{suo2018information, zhan2020susceptible}. In the model, an individual can only take one of the two states, i.e., susceptible (S) or infected (I). An S-state node can be infected by each of its I-state neighbors with infection probability $\beta$. The model is described as follows:

(i) Initially, nodes in the seed set are set to be I-state and the rest nodes are in S-state.

(ii) At each time step $t$, we first find the I-state nodes. For each I-state node $v_i$, we find all the hyperedges $E_i=\{e_{i1}, e_{i2}, \cdots, e_{iq}\}$ that node $v_i$ belongs to. Then a hyperedge $e$ is chosen from $E_i$ uniformly at random. For each of the S-state nodes in $e$, it will be infected by node $v_i$ with infection probability $\beta$.

(iii) We terminate the process until a specific time step $T$ reaches, where $T$ is a control parameter.

\begin{figure}[h!]
\centering		
\includegraphics[scale=1.3]{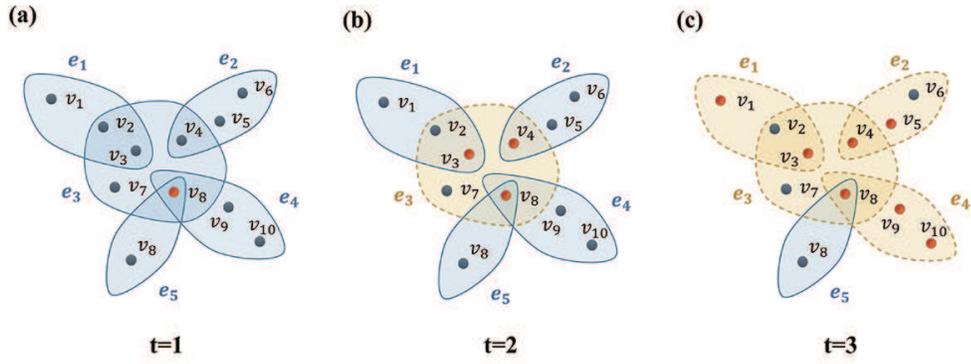}	
\caption{An schematic diagram of the SI spreading model with Contact Process dynamics.}	
\label{fig:SI-CP}		
\end{figure}

We show an illustrative example of the spreading process in Figure~\ref{fig:SI-CP}. At time step $t=1$, node $v_8$ is in I-state. The hyperedge set that contains $v_8$ is $E_3=\{e_3, e_4, e_5\}$. At time step $t=2$, the S-state nodes, i.e., $v_3$ and $v_4$ in hyperedge $e_3$, are infected by node $v_8$. Subsequently, the I-state nodes $v_3$, $v_4$ and $v_8$ infect the S-state nodes in hyperedges $e_1, e_2, e_4$.

\subsection{Degree-based heuristic algorithms}~\label{Degree-based heuristic algorithms}

\begin{figure}[h!]
\centering	
\includegraphics[scale=1.2]{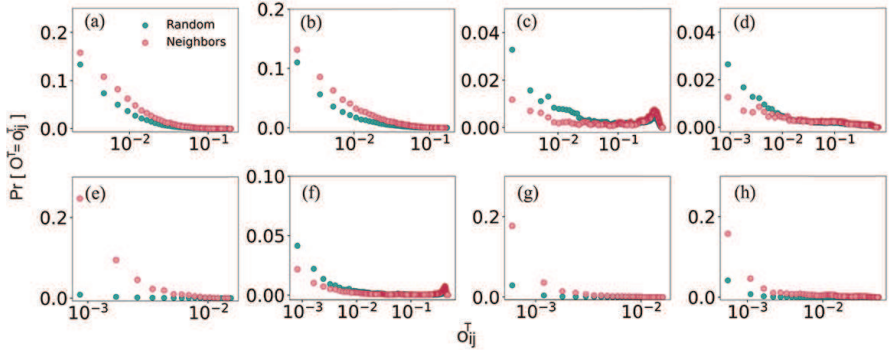}
\caption{The influence overlap distribution of a randomly selected node pair (blue) and a neighboring node pair (pink) in dataset (a) Algebra; (b) Restaurant-Rev; (c) Geometry; (d) Music-Rev; (e) NDC-classes; (f) Bars-Rev; (g) iAF1260b; (h) iJO1366.}
\label{fig:overlap}		
\end{figure}

Given nodes $v_i$ and $v_j$, we suppose that the influenced node sets at time step $T$ by setting node $v_i$ and $v_j$ as the seed node are given by $I_T(v_i)$ and $I_T(v_j)$, respectively. Thus, the influence overlap $o_{ij}^T$ at time step $T$ between $v_i$ and $v_j$ can be defined as $o_{ij}^T=\frac{I_T(v_i)\cap I_T(v_j)}{n}$. In Figure~\ref{fig:overlap}, we show the comparison between the influence overlap distribution of a neighboring node pair and a randomly selected node pair in various hypergraphs. In most of the datasets (i.e., Figure~\ref{fig:overlap}(a), (b), (e), (g) and (h)), the probability that a neighboring node pair have overlapped influence is always higher than that of a randomly selected node pair. It suggests that when we choose one node as the seed, the probability that its neighboring nodes are choosing as the seed should be lower to avoid overlapped influence. Based on this assumption, we propose an adaptive degree-based heuristic algorithm, i.e., Heuristic Degree Discount (HDD), to solve the influence maximization problem.

\textbf{Heuristic Degree Discount (HDD).} In HDD, we aim to punish nodes that have more neighbors in $S$ in each iteration. The details of HDD is shown in Algorithm~\ref{algorithm2}. To conduct the algorithm, we first give the original degree vector of all the nodes as  $deg^{0}=(deg^{0}(1), deg^{0}(2), \cdots, deg^{0}(n))$.

(i) At the initial step, a node $v_i$ that has the largest degree is chosen, i.e., $deg^0(i)=\max\left\{deg^0\right\}$, and added to the seed set $S$. For every neighboring node $v_u$ of $v_i$, we first find the neighbors of $v_u$ in $S$ and collect them as a set $N_S(u)$. Then the adaptive degree of node $v_u$ is updated as $deg^{1}(u)=deg^{0}(u)-Z$, where $Z=|N_S(u)|$ is the number of elements in $N_S(u)$.
For the other nodes that are not the neighbors of $v_i$, e.g., node $v_w$,  $deg^{1}(w)=deg^{0}(w)$. After updating the adaptive degree of every node, we obtain an adaptive degree vector $deg^{1}=(deg^{1}(1), deg^{1}(2), \cdots, deg^{1}(n))$.

(ii) At step $k$, the node that has the largest adaptive degree (denoted as $v_j (v_j \in V \backslash S)$), i.e., $deg^{k-1}(j)=\max\{deg^{k-1}\}$, is chosen,  and we add it to the seed set $S$. For every neighboring node $v_q$ of $v_j$, we first find the neighbors of $v_q$ in $S$ and collect them as a set $N_S(q)$. The adaptive degree of $v_q$ is further updated by $deg^{k}(q)=deg^{k-1}(q)-Z$, where $Z=|N_S(q)|$ is the number of elements in $N_S(q)$. For the other nodes that are not the neighbors of $v_j$, e.g., node $v_w$,  $deg^{k}(w)=deg^{k-1}(w)$.
We obtain a new adaptive degree vector as $deg^{k}=(deg^{k}(1), deg^{k}(2), \cdots, deg^{k}(n))$ after updating the adaptive degree of every node.

(iii) The algorithm is terminated when we obtain $K$ seed nodes.

 We propose a simplified algorithm which considers to give an even penalty for every node in the iteration, i.e., at step $k$, the adaptive degree of $v_q$ is further updated by $deg^{k}(q)=deg^{k-1}(q)-Z$, where $Z=1$. The simplified algorithm is named as Heuristic Single Discount (HSD), which we will use as a baseline for comparison.  We show the details of HSD in Algorithm~\ref{algorithm1}.
\def\SetClass{article}
\begin{algorithm}
\caption{HeuristicDegreeDiscount (HDD)}\label{algorithm2}
  \SetKwInOut{Input}{input}\SetKwInOut{Output}{output}
  \Input{Hypergraph $H(V, E)$ \\ Size of seed nodes $k$}
  \Output{Seed node set $S$}
 \textbf{Initialization:} $deg^{0} \leftarrow$ Degree of each node;\\

  \While{$|S|\leq K$}
    {$k \leftarrow |S|$\\
    $ v_j(v_j \in V \backslash S)  \leftarrow max\left\{deg^{k}(j)\right\}$\\
    $S \leftarrow S   \cup \left\{ v \right\}$\\
    $ N(j)  \leftarrow \text{Neighbors of node} \; v_j$\\

    \For {$v_q  \text{ in }  N(j) $}{
        $ N_s(q) \leftarrow \text{the neighbors of } v_q \text{ in } S $\\
        $ deg^{k}(q)$ = $deg^{k-1}(q) - |N_s(q)|$
    }
    \For {$v_w  \text{ not in }  N(j) $}{
        $ deg^{k}(w)=deg^{k-1}(w)$
    }
    }
\end{algorithm}

\begin{algorithm}

\caption{HeuristicSingleDiscount (HSD) }\label{algorithm1}
  \SetKwInOut{Input}{input}\SetKwInOut{Output}{output}
  \Input{Hypergraph $H(V, E)$ \\ Size of seed nodes $K$}
  \Output{Seed node set $S$}
   \textbf{Initialization:} $deg^{0} \leftarrow$ Degree of each node;\\

  \While{$|S|\leq K$}
    {$k \leftarrow |S|$\\
    % $ Sort(deg^{k}) \leftarrow \text{Ranking the degree based vector }deg^{k} \text{;}$\\
    % $ v_i(v_i \in V \backslash S)  \leftarrow \text{pop from the top in} \; Sort(deg^{k}) $\\
    $ v_j(v_j \in V \backslash S)  \leftarrow \max\left\{deg^{k}(j)\right\}$\\
    $S \leftarrow S   \cup \left\{ v_j \right\}$\\
    $ N(j)  \leftarrow \text{Neighbors of node} \; v_j$\\
    % \For{$v_q \text{ in } N(v_j)$}{
    % $deg^{k+1}(v_q) = deg^{k}(v_j) - 1$\\
    % }
    \For {$v_q  \text{ in }  N(j) $}{
        $ N_s(q) \leftarrow \text{the neighbors of } v_q \text{ in } S $\\
        $ deg^{k}(q)$ = $deg^{k-1}(q) - 1$
    }
    \For {$v_w  \text{ not in }  N(j) $}{
        $ deg^{k}(w)=deg^{k-1}(w)$
    }

    }
\end{algorithm}

\subsection{Baselines and Extended Algorithms}
To verify the performance of our algorithms, we propose two algorithms extended from ordinary network, i.e., H-RIS and H-CI, and choose state-of-the-art algorithms, i.e., Greedy, Hyperdegree, Degree, proposed by other researchers as baselines. The details of each algorithm are given as follows.
% \begin{itemize}

\textbf{Hyper Reverse Influence Sampling (H-RIS)}. Reverse Influence Sampling (RIS) algorithm was proposed to solve the influence maximization problem in an ordinary network~\cite{borgs2014maximizing}.  In this work, we propose to extend the reverse influence sampling algorithm to a hypergraph by first introducing the following two definitions:

    \begin{myDef}
    \label{(HYPER REVERSE REACHABLE SET)}(HYPER REVERSE REACHABLE SET). Given a hypergraph $H(V, E)$, we remove each hyperedge from it with probability $1-\beta$ and obtain a sub-hypergraph $H'(V', E')$. Given a node $v\in H$, we define the hyper reverse reachable (HRR) node set that can reach node $v$ in $H'$.
    \end{myDef}
    \begin{myDef}
    \label{(RANDOM RR SET)}(RANDOM HRR SET). For a randomly selected node $v \in H$, a random HRR set is defined as a HRR set which is randomly sampled from the pruned hypergraph $H'$.
    \end{myDef}

    Based on the above definitions, we illustrate the details of the H-RIS algorithm in Algorithm~\ref{algorithm3}, which mainly contains the following two steps:

    (i) We generate $\eta$ random HRR sets, in which $\eta$ is a tunable parameter.

    (ii) In each round of seed selection, we add node $v_q$ with the highest frequency in the generated HRR sets to the seed set $S$. Then, the HRR sets that contain node $v_q$ are removed. The selection rounds will be terminated until $K$ seed nodes are selected.
\begin{algorithm}
\caption{Hyper Reverse Influence Sampling (H-RIS) }\label{algorithm3}
  \SetKwInOut{Input}{input}\SetKwInOut{Output}{output}
  \Input{Hypergraph $H(V, E)$ \\ Size of seed nodes $k$ \\ infection probability $\beta$}
  \Output{Seed node set $S$}
%   Compute the hyperdegree(all their neighbors) of all nodes.;\\
  Initialize $S=\emptyset$, $U=\emptyset$. U is a set of $HRR$.\\
  $H'(V', E') \leftarrow \text{Remove hyperedges with probability} \;1-\beta\; \text{ from hypergraph } H(V, E)$\\
  \For{$i=1\; \text{to}\; \eta $}{

    $v_i \leftarrow \text{Select the node randomly}$\\
    $HRR \leftarrow \text{Aquire nodes reachable to } v_i \;\text{ from }\; H'(V', E')$\\
    $U \leftarrow U \cup \left\{ HRR \right\}$
  }
  \While{$|S|\leq K$}{
    $v_q \leftarrow \text{Node with the highest frequency in} \;U$ \\
    $S \leftarrow S \cup \left\{ v_q \right\}$\\
    $\text{Delete the } HRR\; \text{containing}\; v_q\; \text{from}\; U$
  }
  \end{algorithm}

 The algorithm suggests that if a node appears more frequently in different HRR sets, it will have a higher probability to influence the other nodes. Correspondingly, the more HRR sets that the seed set $S$ covers, the more likely that $S$ will have a large expected influence. We set $\eta=200$ to conduct the experiments.

\textbf{Hyper Collective Influence (H-CI)}.
Collective Influence (CI) was first proposed to select seed nodes based on the degree of distant nodes in an ordinary network \cite{morone2015influence} . We extend the algorithm to a hypergraph by simply replacing node degree by the hyperdegree in the definition of hyper collective influence. A ball $Ball(v_i, l)$ is defined as a set of nodes which contains nodes inside a ball of radius $l$, where $l$ denotes as the shortest path from a node in $Ball(v_i, l)$ to node $v_i$. $\partial Ball(v_i, l)$ is the frontier of $Ball(v_i, l)$, i.e., the path length of any node inside $\partial Ball(v_i, l)$ to node $v_i$ equals to $l$. We define the HCI of node $v_i$, which is read as

%  \begin{myDef}
%     \label{(HYPER COLLECTIVE INFLUENCE)}(HYPER COLLECTIVE INFLUENCE). The hyper collective influence $HCI_l(i)$ of node $v_i$ satisfies the following equation:
\begin{equation}
    HCI_l(i) = (d^H(i)-1)\sum_{v_j \in \partial Ball(v_i, l)}(d^H(j)-1),
\end{equation}
    % \end{myDef}
where $d^H(i)$ is the hyperdegree of node $v_i$.

 Given a specific value of $l$, we calculate the HCI of every node in the hypergraph and choose top $K$ nodes that have the highest HCI value as the seeds for influence maximization problem.  In this work, the tunable parameter $l$ is set as $1$ and $2$, and we name the algorithms as H-CI$(l=1)$ and H-CI$(l=2)$, respectively.

\textbf{Greedy.}
Greedy algorithm gives a guaranteed approximation of influence spread by accurately approximating influence spread with high computational complexity. The algorithm can be extended to a hypergraph~\cite{kempe2003maximizing}, which is shown in Algorithm~\ref{algorithm4}.
We denote $S_{k-1}$ as the seed nodes that are selected at round $k-1$, the expected influence spread by $S_{k-1}$ is given  by $\sigma(S_{k-1})$. The marginal influence spread by adding node $v$ to the seed set at round $k$ is given by $\sigma(S_{k-1} \cup \left\{v\right\})-\sigma(S_{k-1})$. At the beginning of the algorithm, $S$ is set to be empty. At round $k$, we calculate the expected influence spread $\sigma(S_{k-1} \cup \left\{v\right\})$ for each $v$, where $v\in V\backslash S_{k-1}$. Node $v_k$ with the largest marginal influence contribution ($v_k = \arg \max_{v\notin S_{k-1}}\left\{
    \sigma(S_{k-1} \cup \left\{v\right\})-\sigma(S_{k-1} \right\}$) is added to the seed set , i.e., $S_{k}=S_{k-1}\cup \left\{v_k\right\}$.
The algorithm is terminated until $K$ seed nodes are selected.

\begin{algorithm}
\caption{Greedy}\label{algorithm4}
\SetKwInOut{Input}{Input}\SetKwInOut{Output}{Output}
\Input{Hypergraph $H$ \\ Size of seed nodes $K$}
\Output{Seed node set $S$}
Initialize $S=\emptyset$, $k = 0$.\\
\While{$|S|\leq K$}{
% \For{$v \leftarrow v \in V \backslash S$}{
%     $\sigma(v) = \sigma(S \cup \left\{v\right\}) $
% }
$v_k = \arg \max_{v\notin S_{k-1}}\left\{ \sigma(S_{k-1} \cup \left\{v\right\}-\sigma(S_{k-1})  \right\}$ \\
$ S_{k} = S_{k-1} \cup  \left\{ v_k\right\}$ \\
$ k =  k+1 $
}
\end{algorithm}

\textbf{HyperDegree.}
We compute the hyperdegree of every node in a hypergraph and choose top $K$ nodes with the highest hyperdegree as the seeds for influence maximization problem.

\textbf{Degree.} Similar to HyperDegree, we calculate the degree of every node in a hypergraph and select top $K$ nodes with the highest degree into the seed set $S$ for influence maximization problem.

\section*{4. Results}\label{EXPERIMENTS}
In this section, extensive experiments on eight hypergraphs generated by real-world data are carried out to validate the effectiveness and efficiency of the proposed algorithms. Besides, we also test the robustness of our algorithms on synthetic hypergraphs generated by different degree heterogeneities.

\subsection{Experimental Evaluation on Real Data}
To compare the performance of different algorithms on solving the influence maximization problem, we use the seed set obtained by each algorithm as the seed nodes for the SI spreading model with contact process running on various hypergraphs. In the SI spreading model, we show the results of different combinations of infection probability $\beta$ and the termination step.
The expected influence spread by seed set $S$ is given by the average of the outbreak sizes over $500$ realizations for each algorithm. In addition, the size of the seed set $S$ varies from $1$ to $25$ in our experiments. All the algorithms are written in Python language, and each of them independently runs on a sever with 2.20GHz Intel(R) Xeon(R) Silver 4114 CPU and 90G memory.

\begin{figure}[h!]
\centering	
\includegraphics[scale=1.2]{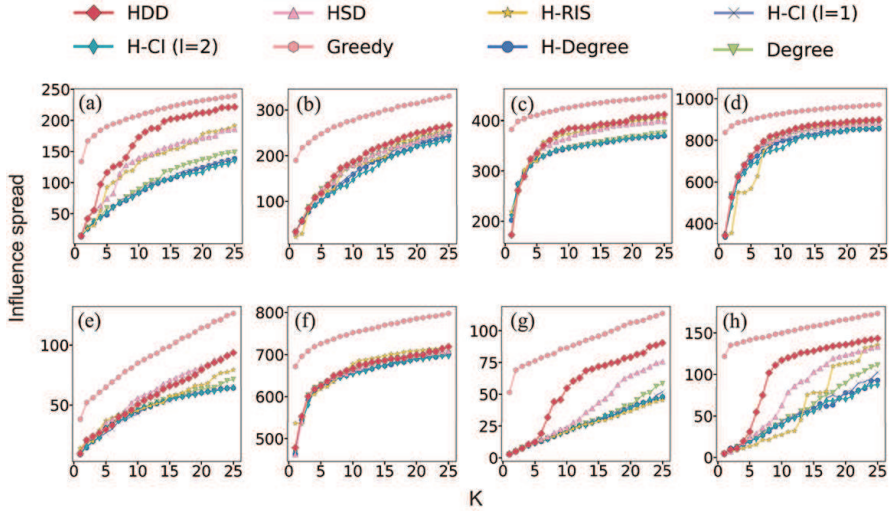}	
\caption{Expected influence spread as a function of the seed set size $K$ for each algorithm in hypergraphs: (a) Algebra; (b) Restaurant-Rev; (c) Geometry; (d) Music-Rev; (e) NDC-classes; (f) Bars-Rev; (g) iAF1260b; (h) iJO1366. We set $\beta=0.01, T=25$.}
\label{fig:Agr_pfm}		
\end{figure}

The influence spread results of different algorithms when $\beta=0.01, T=25$ are given in Figure~\ref{fig:Agr_pfm}, Table~\ref{AUC} and~\ref{algorithm_running_time}. In Figure~\ref{fig:Agr_pfm}, we depict the expected influence spread as a function of the seed set size $K$, the area under each of the influence spread curve (AUC) is further given in Table~\ref{AUC}. The best performance is obtained by Greedy algorithm, which comprehensively considers the topological and dynamical information. The algorithms (i.e., HDD and HSD) we proposed perform second best in almost all the hypergraphs, except for hypergraph \textbf{Bars-Rev} with AUC slightly lower than H-RIS (i.e., $0.15\%$ lower than H-RIS). As it is illustrated in Section~\ref{Degree-based heuristic algorithms}, the basic assumption for HDD and HSD is that when we choose one node as the seed, the
probability that its neighboring nodes are choosing as the seed should be lower to avoid overlapped influence. HDD, HSD and Degree are algorithms based on the node degree, but HDD, HSD perform much better than Degree algorithm in all the hypergraphs. In hypergraphs such as \textbf{Algebra}, \textbf{Restaurant-Rev}, \textbf{NDC-classes}, \textbf{ iAF1260b} and \textbf{iJO1366}, the probability that a neighboring node pair have overlapped
influence is higher than that of a randomly selected node pair (Figure~\ref{fig:overlap}). Accordingly, the AUC values in these hypergraphs derived from HDD, HSD are also relatively larger than other algorithms except Greedy, which is shown in Table~\ref{AUC}. It suggests that the assumption of reducing influence overlap can help to improve the performance of influence maximization algorithms. The fact that HDD is superior to HSD in finding seed nodes further implies that considering an uneven penalty for each node in the design of the algorithm is more reasonable for influence maximization. H-CI($l=1$), H-CI($l=2$) and H-Degree are algorithms based on the hyperdegrees of the nodes, and we find that H-CI($l=1$) and H-CI($l=2$) perform slightly better than H-Degree. It indicates that considering the hyperdegree of distant nodes can help to improve the selection of seeding nodes. The AUC values of the other combinations of $\beta$ and $T$ are given in Table~\ref{AUC_beta_T_1}, ~\ref{AUC_beta_T_2}, ~\ref{AUC_beta_T_3}, respectively, which is consistent with the results we obtained from $\beta=0.01, T=25$.

We further show the time cost for singling out seed node set ($\beta=0.01, T=25$) in Table~\ref{algorithm_running_time}, where the time cost is the average over 10 realizations for each algorithm. We set $K=25$. Even though Greedy algorithm performs the best for influence spread, it has the highest time cost, i.e., it takes a few hours or days for each realization. Besides, H-RIS and HCI($l=2$) also have high computational complexity compared to the remaining algorithms. H-Degree and Degree take the least time cost but with low AUC. In contrast, HDD and HSD can achieve relatively high AUC with low time cost (within 50 seconds) in all the hypergraphs.

% \begin{figure}[h!]
% \centering	
% \includegraphics[scale=0.43]{pfm_2.eps}	
% \caption{Expected influence spread as a function of the seed set size $K$ for each algorithm in hypergraphs: (a) Algebra; (b) Restaurant-Rev; (c) Geometry; (d) Music-Rev; (e) NDC-classes; (f) Bars-Rev; (g) iAF1260b; (h) iJO1366. We set $\beta=0.01, T=25$.}
% \label{fig:Agr_pfm}		
% \end{figure}

\begin{table}[htb]
		\centering
		\caption{AUC scores obtained by each of the curves shown in Figure~\ref{fig:Agr_pfm} for algorithms, i.e., HDD, HSD, H-RIS, H-CI($l=1$), H-CI($l=2$), H-degree and Degree. The best performance, i.e., the largest AUC score, is shown by $**$ and the second best is shown by $*$ in each hypergraph. We set $\beta=0.01, T=25$.}
		\resizebox{\textwidth}{!}{
		\begin{tabular}{|l|c|c|c|c|c|c|c|}\hline
			hypergraph&\multicolumn{7}{c|}{AUC}\\\hline
			&\textbf{HDD}&HSD&H-RIS&H-CI($l=1$)&H-CI($l=2$)&H-degree&Degree\\
			Algebra&\textbf{0.2068}$^{**}$&0.1668$^*$&0.1630&0.1155&0.1114&0.1132&0.1232\\
			Restaurant-Rev&\textbf{0.1549}$^{**}$&0.1474$^*$&0.1453&0.1384&0.1315&0.1355&0.1471\\
			Geometry&\textbf{0.1501}$^{**}$&0.1468&0.1487$^*$&0.1385&0.1386&0.1385&0.1388\\
			Music-Rev&\textbf{0.1474}$^{**}$&0.1453$^*$&0.1407&0.1423&0.1386&0.1407&0.1449\\
			NDC-classes&0.1601$^*$&\textbf{0.1691}$^{**}$&0.1438&0.1292&0.1311&0.1309&0.1359\\
            Bars-Rev&0.1439&0.1433$^*$&\textbf{0.1448}$^{**}$&0.1420&0.1414&0.1416&0.1430\\
            iAF1260b&\textbf{0.2452}$^{**}$&0.1680$^*$&0.1115&0.1184&0.1162&0.1164&0.1243\\
            iJO1366&\textbf{0.2269}$^{**}$&0.1751$^*$&0.1359&0.1166&0.1084&0.1109&0.1262\\\hline
		\end{tabular}
		}
		\label{AUC}
	\end{table}

\begin{table}[htb]
		\centering
		\caption{AUC scores obtained by our algorithms and baselines. The best performance, i.e., the largest AUC score, is shown by $**$ and the second best is shown by $*$ in each hypergraph. We set $\beta=0.005, T=35$.}
		\resizebox{\textwidth}{!}{
		\begin{tabular}{|l|c|c|c|c|c|c|c|}\hline
			hypergraph&\multicolumn{7}{c|}{AUC}\\\hline
			&\textbf{HDD}&HSD&H-RIS&H-CI($l=1$)&H-CI($l=2$)&H-degree&Degree\\
			Algebra&\textbf{0.2335}$^{**}$&0.1702$^*$&0.1669&0.1067&0.1022&0.1048&0.1156\\
            Restaurant-Rev&\textbf{0.1606}$^{**}$&0.1476&0.1528$^*$&0.1358&0.1255&0.1306&0.1471\\
            Geometry&\textbf{0.1576}$^{**}$&0.1507&0.1514$^*$&0.1351&0.1351&0.1344&0.1357\\
            Music-Rev&0.1554$^*$&0.1475&\textbf{0.1562}$^{**}$&0.1352&0.1279&0.1317&0.1462\\
            NDC-classes&0.1547$^*$&\textbf{0.1674}$^{**}$&0.1315&0.1341&0.1356&0.1359&0.1408\\
            Bars-Rev&\textbf{0.1472}$^{**}$&0.1470$^*$&0.1258&0.1453&0.1437&0.1444&0.1467\\
            iAF1260b&\textbf{0.2306}$^{**}$&0.1605$^*$&0.1304&0.1190&0.1176&0.1181&0.1237\\
            iJO1366&\textbf{0.2696}$^{**}$&0.1890$^*$&0.0920&0.1121&0.1064&0.1067&0.1243\\\hline
		\end{tabular}
		}
		\label{AUC_beta_T_1}
	\end{table}
	
\begin{table}[htb]
		\centering
        \caption{AUC scores obtained by our algorithms and baselines. The best performance, i.e., largest AUC score, is shown by $**$ and the second best is shown by $*$ in each hypergraph. We set $\beta=0.015, T=15$.}
        \resizebox{\textwidth}{!}{
		\begin{tabular}{|l|c|c|c|c|c|c|c|}\hline
			hypergraph&\multicolumn{7}{c|}{AUC}\\\hline
			&\textbf{HDD}&HSD&H-RIS&H-CI($l=1$)&H-CI($l=2$)&H-degree&Degree\\
			Algebra&\textbf{0.2235}$^{**}$&0.1717$^*$&0.1498&0.1132&0.1089&0.1107&0.1222\\
            Restaurant-Rev&0.1562$^*$&0.1461&\textbf{0.1586}$^{**}$&0.1356&0.1266&0.1315&0.1455\\
            Geometry&\textbf{0.1536}$^{**}$&0.1488&0.1496$^*$&0.1367&0.1370&0.1368&0.1375\\
            Music-Rev&0.1497$^*$&0.1451&\textbf{0.1585}$^{**}$&0.1374&0.1312&0.1343&0.1438\\
            NDC-classes&0.1645$^*$&\textbf{0.1760}$^{**}$&0.1025&0.1368&0.1382&0.1385&0.1435\\
            Bars-Rev&\textbf{0.1441}$^{**}$&0.1440$^*$&0.1437&0.1422&0.1411&0.1415&0.1433\\
            iAF1260b&\textbf{0.2407}$^{**}$&0.1652$^*$&0.1193&0.1185&0.1160&0.1165&0.1238\\
            iJO1366&\textbf{0.2313}$^{**}$&0.1693&0.1815$^*$&0.1048&0.0986&0.0995&0.1149\\\hline
		\end{tabular}
		}
		\label{AUC_beta_T_2}
	\end{table}

\begin{table}[htb]
		\centering
        \caption{AUC scores obtained by our algorithms and baselines. The best performance, i.e., largest AUC score, is shown by $**$ and the second best is shown by $*$ in each hypergraph. We set $\beta=0.02, T=10$.}
        \resizebox{\textwidth}{!}{
		\begin{tabular}{|l|c|c|c|c|c|c|c|}\hline
			hypergraph&\multicolumn{7}{c|}{AUC}\\\hline
			&\textbf{HDD}&HSD&H-RIS&H-CI($l=1$)&H-CI($l=2$)&H-degree&Degree\\
			Algebra&\textbf{0.2339}$^{**}$&0.1714$^*$&0.1582&0.1088&0.1033&0.1069&0.1175\\
            Restaurant-Rev&0.1582$^*$&0.1454&\textbf{0.1657}$^{**}$&0.1335&0.1233&0.1291&0.1448\\
            Geometry&\textbf{0.1600}$^{**}$&0.1526$^*$&0.1513&0.1336&0.1339&0.1336&0.1350\\
            Music-Rev&\textbf{0.1593}$^{**}$&0.1486&0.1568$^*$&0.1335&0.1256&0.1296&0.1466\\
            NDC-classes&0.1571$^*$&\textbf{0.1688}$^{**}$&0.1257&0.1346&0.1361&0.1365&0.1412\\
            Bars-Rev&0.1449$^*$&\textbf{0.1451}$^{**}$&0.1395&0.1432&0.1408&0.1416&0.1448\\
            iAF1260b&\textbf{0.2361}$^{**}$&0.1634$^*$&0.1203&0.1197&0.1178&0.1181&0.1245\\
            iJO1366&\textbf{0.2684}$^{**}$&0.1873$^*$&0.1004&0.1104&0.1048&0.1064&0.1223\\ \hline
		\end{tabular}
		}
		\label{AUC_beta_T_3}
	\end{table}

\begin{table}[htb]
		\centering
		\caption{Time cost for each algorithm. The running time are given by the average over $10$ realizations, the seed set size is set as $K=25$. We set $\beta=0.01, T=25$. }
			\resizebox{\textwidth}{!}{
			\begin{tabular}{|l|c|c|c|c|c|c|c|c|}\hline
			hypergraph&\multicolumn{8}{c|}{Time cost (seconds)}\\\hline
		    &\textbf{HDD}&HSD&H-RIS&H-CI($l=1$)&H-CI($l=2$)& Greedy&H-degree&Degree\\
			Algebra&19.2191&1.7816&101.4887&1.4566&489.7527&15991.0652&0.0290&0.0216\\
			Restaurant-Rev&9.9571&1.5490&79.2794&1.1639&219.7846&23630.6722&0.0434&0.0311\\
			Geometry&46.6095&2.6269&173.1826&2.6695&2369.7399&53966.0565&0.0293&0.0294 \\
            Music-Rev &30.4322&3.4626&618.9846&3.6286&2164.4441&144976.2404&0.0623&0.0653\\
            NDC-classes &8.7360&2.9378&4317.1805&1.3690&5748.8244&18891.5252&0.0560&0.0637\\
            Bar-Rev &30.9617&3.6873&3472.9475&3.9713&12715.2541&131718.2580&0.0780&0.0621\\
            iAF1260b &14.8016&4.1104&3532.0354&1.9957&92402.2618&15396.0684&0.1050&0.0885\\
            iJO1366 &19.3894&4.5724&9123.7571&2.2732&91108.2699&30233.7775&0.0824&0.0943\\\hline
		\end{tabular}
		}
		\label{algorithm_running_time}
	\end{table}

\clearpage
\subsection{Experimental Evaluation on Synthetic Hypergraphs}
\begin{figure}[htb]
\centering
\includegraphics[scale=1.2]{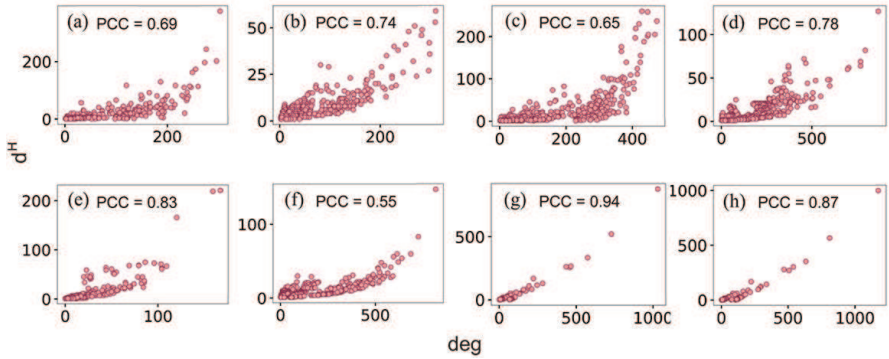}	
\caption{The correlation between node degree and hyperdegree in hypergraphs generated by real-world datasets: (a) Algebra; (b) Restaurant-Rev; (c) Geometry; (d) Music-Rev; (e) NDC-classes; (f) Bars-Rev; (g) iAF1260b; (h) iJO1366. In each figure, we show the Pearson correlation coefficient (PCC) between node degree and hyperdegree in each of the hypergraphs.}
\label{fig:correlation-real}
\end{figure}

The influence maximization methods we proposed, i.e., HDD and HSD, are degree-based heuristic methods. To check the robustness of our method over the change of the degree heterogeneity~\cite{st2022influential}, we evaluate the performance of our methods on synthetic hypergraphs.
Figure~\ref{fig:correlation-real} indicates that the node degree is positively correlated with the corresponding hyperdegree in the hypergraphs generated by real data, with the Pearson Correlation Coefficient (PCC) higher than 0.5.
Therefore, we choose to use
HyperCL~\cite{lee2021hyperedges}, which is a random hypergraph generator, to generate hypergraph with a certain hyperdegree distribution. The details of HyperCL are given as follows:

Initially, we suppose the hyperdegree and the hyperedge size sequence of a hypergraph  $H(V, E)$ are given as $ \left\{d^H(1), d^H(2),  \cdots,d^H(n)\right\}$ and \linebreak $ \left\{d^E(1), d^E(2), \cdots,d^E(m)\right\}$, respectively. For each $e_i \in E$, the nodes belong to $e_i$ are sampled independently. That is to say, we select node $v_j$ into $e_i$ with probability proportional to  its hyperdegree (i.e., the probability is $\frac{d^H(j)}{\sum_{j=1}^{n}d^H(j)})$ until the size of the hyperedge $e_i$ reaches $d^E(e_i)$. Specifically, duplicated nodes are ignored in each hyperedge generation. The algorithm is terminated until the size of each hyperedge reaches the pre-set size.

\begin{figure}[htb]
\centering
\includegraphics[scale=1]{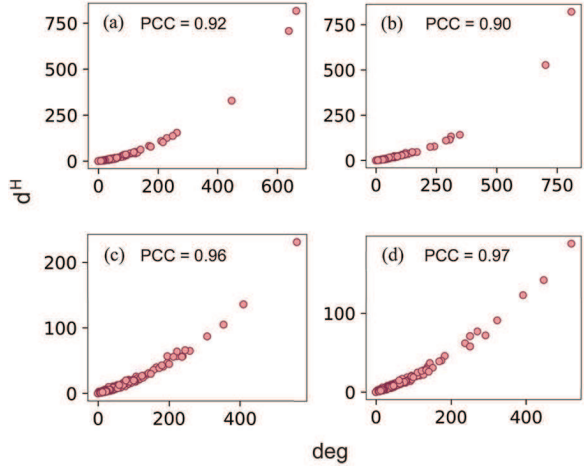}
\caption{The correlation between node degree and hyperdegree in synthetic hypergraphs generated by HyperCL via different exponents: (a) $\Theta=2$; (b) $\Theta=2.1$; (c) $\Theta=2.3$; (d)
$\Theta=2.5$. In each figure, we show the Pearson correlation coefficient (PCC) between node degree and hyperdegree in each of the hypergraphs. The number of nodes and hyperedges in the hypergraphs are set as $n=1000$ and $m=1000$, respectively.}
\label{fig:correlation_hyperCL}
\end{figure}

\begin{figure}[htb]
\centering
\includegraphics[scale=1]{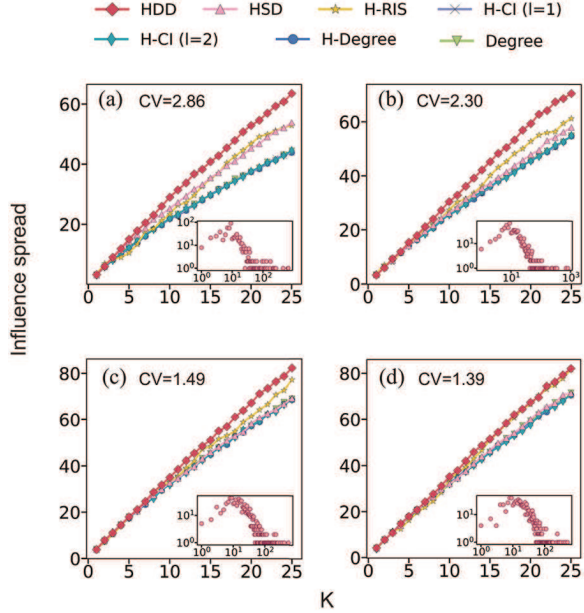}	
\caption{Expected influence spread as a function of $K$ for each algorithm on synthetic hypergraphs generated by HyperCL with  (a) $\Theta=2$; (b) $\Theta=2.1$; (c) $\Theta=2.3$; (d)$\Theta=2.5$. The inset in each figure depicts the degree distribution of each hypergraph. The number of nodes and hyperedges in the hypergraphs are set as $n=1000$ and $m=1000$, respectively.}	
\label{fig:hypercl-AUC}		
\end{figure}

In the HyperCL, the hyperdegree sequence is generated by a hyperdegree distribution $p(d^H) \sim (d^H)^{-\Theta}$, where the exponent $\Theta$ is a tunable parameter.
As the value of exponent $\Theta$ increases, the hyperdegree distribution would change from heterogeneous to homogeneous. In this work, the exponent value is set as $\Theta = 2, 2.1, 2.3$ and $2.5$.  The hyperedge size sequence generated by a uniform distribution with the minimum and maximal size setting as $0$ and $10$, respectively. We use
coefficient of variation (CV), which is defined as the ratio of the standard deviation to the mean, to measure the heterogeneity of the degree distribution of a synthetic hypergraph.  In Figure~\ref{fig:hypercl-AUC}, we show that as $\Theta$ increases, the CV decreases, indicating that the degree distribution would be more homogeneous.  We further show the correlation between the degree and hyperdegree of a node in the synthetic hypergraphs generated by HyperCL in Figure~\ref{fig:correlation_hyperCL}, where the PCC is
higher than 0.9 in hypergraphs generated by different  hyperdegree distribution.

We show the performance of our methods and the baselines on synthetic hypergraphs on influence maximization problem  in Figure~\ref{fig:hypercl-AUC} and Table~\ref{AUC syn}, respectively. We observe that HDD outperforms all the other methods in all the hypergraphs and H-RIS performs the second best. In addition, as $\Theta$ decreases, i.e., the degree distribution is more heterogeneous, HDD can gain more improvement in AUC than H-RIS (Table~\ref{AUC syn}). It suggests that HDD tends to be more suitable for solving influence maximization problem in hypergraphs with heterogeneous degree distribution, which is
common in real world.

% \clearpage

\begin{table}[htb]
		\centering
		\caption{AUC scores obtained by each of the curve presented in Figure~\ref{fig:hypercl-AUC} for algorithms, i.e., HDD,  HSD, H-RIS, H-CI($l=1$), H-CI($l=2$), H-degree and Degree. The best performance, i.e., the largest AUC score, is shown by $**$ and the second best is shown by $*$ in each hypergraph.}
		\resizebox{\textwidth}{!}{
		\begin{tabular}{|l|c|c|c|c|c|c|c|l|l|}\hline
			Hypergraph&\multicolumn{7}{c|}{AUC}&CV&Gain\\\hline
			&\textbf{HDD}&HSD&H-RIS&H-CI($l=1$)&H-CI($l=2$)&H-degree&Degree& & \\
% 			$\hat{H}(\Theta=5)$&0.1473&0.1482$^{**}$&0.1371$^*$&0.1415&0.1392&0.1340&0.1468&8.10\%\\
% 			$\hat{H}(\Theta=3)$&0.1424&0.1458$^{**}$&0.1470$^*$&0.1415&0.1405&0.1407&0.1420&-0.82\%\\
            $\hat{H}(\Theta=2)$&\textbf{0.1766}$^{**}$&0.1540$^*$&0.1510&0.1293&0.1296&0.1293&0.1300&2.86&14.68\%\\
            $\hat{H}(\Theta=2.1)$&\textbf{0.1686}$^{**}$&0.1411&0.1497$^*$&0.1349&0.1349&0.1349&0.1357&2.30&12.63\%\\
            $\hat{H}(\Theta=2.3)$&\textbf{0.1581}$^{**}$&0.1398&0.1476$^*$&0.1387&0.1383&0.1382&0.1394&1.49&7.11\%\\
            $\hat{H}(\Theta=2.5)$&\textbf{0.1553}$^{**}$&0.1405&0.1510$^*$&0.1384&0.1375&0.1384&0.1388&1.39&2.85\%\\
            \hline
		\end{tabular}
		}
		\label{AUC syn}
	\end{table}

\section*{5. Conclusions}\label{Conclusions}
Much effort has been devoted to find influential node set in ordinary networks. In this work, we tackle the challenge on influence maximization problem in hypergraphs, which aims to identify $K$ initial spreaders that can maximize the expected outbreak size of a certain spreading dynamics. We start with a simple spreading model, i.e., susceptible-infected (SI) model with contact process dynamics. Based on the fact that the influence overlap between a node and its neighboring nodes is usually high in hypergraphs generated by real data, we propose an algorithm called Heuristic Degree Discount (HDD) to solve the influence maximization problem in hypergraphs. The algorithm iteratively gives large penalty to nodes that have more neighbors in the existing seed set and thus these nodes are less likely to be chosen as seeds. To validate the effectiveness of our algorithm, we demonstrate a list of baseline algorithms, including the ones proposed by previous researchers as well as algorithms extended from ordinary networks. We perform the experiments on eight hypergraphs generated by real data from various domains. The results show that HDD is superior to the baselines in
terms of  accuracy (except Greedy) almost in all the hypergraphs with different infection probability. In addition, our algorithm also shows good performance in terms of efficiency. As HDD is based on the node degree, we further test the performance on synthetic hypergraphs generated by HyperCL, which can generate hypergraphs with different hyperdegrees. The results demonstrate HDD gains more AUC scores in hypergraphs with high degree heterogeneity.

Heuristic algorithms have been widely utilized to solve the influence maximization on ordinary networks due to its low computational complexity. In this work, we confine to use a simple heuristic from hypergraph, i.e., degree, to design algorithm for identifying seed node set, which shows high performance. We deem that more high-order properties from hypergraph could be used for influence maximization. Moreover, our algorithm framework could also be promising in solve influence maximization problem for other dynamic processes, such as threshold model~\cite{xu2022dynamics}, independent cascade model~\cite{ma2022hyper} and other epidemic models~\cite{jhun2019simplicial, de2020social}.

\section*{Acknowledgements}

This work was supported by Natural Science Foundation of Zhejiang Province (Grant Nos. LQ22F030008 and LR18A050001), the National Natural Science Foundation of China (Grant Nos. 92146001, 61873080 and 61673151), the Major Project of The National Social Science Fund of China (Grant No. 19ZDA324), and the Scientific Research Foundation for Scholars of HZNU (2021QDL030).

\section*{Author Contributions Statement}
\
All authors planed the study. All authors performed the experiments and prepared the figures. All authors analyzed the results and wrote the manuscript. All authors read and approved the final manuscript.

\section*{Data and Code Availability}
The data and code used in this work can be access via: https://github.com/DDMXIE/Influence-maximization-on-hypergraphs.

\bibliography{TempExample.bib}

% \section*{Acknowledgements (not compulsory)}

% Acknowledgements should be brief, and should not include thanks to anonymous referees and editors, or effusive comments. Grant or contribution numbers may be acknowledged.

% \section*{Author contributions statement}

% Must include all authors, identified by initials, for example:
% A.A. conceived the experiment(s),  A.A. and B.A. conducted the experiment(s), C.A. and D.A. analysed the results.  All authors reviewed the manuscript.

% \section*{Additional information}

% To include, in this order: \textbf{Accession codes} (where applicable); \textbf{Competing interests} (mandatory statement).

% The corresponding author is responsible for submitting a \href{http://www.nature.com/srep/policies/index.html#competing}{competing interests statement} on behalf of all authors of the paper. This statement must be included in the submitted article file.

\end{document}